\documentclass[aip, apl, reprint]{revtex4-1}

\draft % marks overfull lines with a black rule on the right
\usepackage{graphicx}% Include figure files
\usepackage{dcolumn}% Align table columns on decimal point
\usepackage{bm}

\begin{document}

%\preprint{}

\title{Bulk-sensitive Imaging of Twin Domains in La$_{2-x}$Sr$_x$CuO$_4$ under Uniaxial Pressure}



\author{Xin Yu Zheng}
%\thanks{}
\affiliation{Department of Physics, University of Toronto, 60 St.  George Street, Toronto, ON M5S 1A7, Canada}

\author{Renfei Feng}
\affiliation{Canadian Light Source, 44 Innovation Blvd., Saskatoon, SK S7N 2V3, Canada}

\author{D. S. Ellis}
\affiliation{Department of Physics, University of Toronto,
60 St.  George Street, Toronto, ON M5S 1A7, Canada}
\altaffiliation{Department of Materials Science and Engineering, Technion Israel Institute of Technology, Technion City, Haifa 3200003, Israel}

\author{Young-June Kim}
\affiliation{Department of Physics, University of Toronto,
60 St.  George Street, Toronto, ON M5S 1A7, Canada}


\date{\today}

\begin{abstract}
We report our study of twin domains in $La_{2-x}Sr_x CuO_4$ under uniaxial pressure. Using bulk-sensitive x-ray microdiffraction in Laue geometry, we image the distribution of twin domains at room temperature. When compressive uniaxial pressure is applied along one of the in-plane crystallographic axes, the domain population changes dramatically. We observe that the twin domain with {\em shorter} lattice parameter along the direction of pressure is unstable under compression, and disappears completely with only moderate pressure. On the other hand, application of tensile pressure changes the domain structure only slightly, demonstrating the asymmetric response of the sample to uniaxial pressure. Our observations suggest that a crystal's response to uniaxial pressure is complex and could deviate easily from the linear-response regime.
\end{abstract}


\maketitle %\maketitle must follow title, authors, abstract and \pacs


Understanding of stress and strain has been an integral part of materials science and engineering. Mechanical properties of bulk engineering materials are influenced significantly by the residual stress, while epitaxial strain plays an important role in determining structural and electronic properties of many thin film materials. Unlike such naturally occurring, unavoidable stress/strain, a deliberate application of uniaxial pressure\footnote{We use the term pressure in our experiment, since both stress and strain are second-rank tensor quantities, and thus not easy to define uniaxial direction.} to perturb a crystal is gaining interest as an experimental tool in condensed matter physics.\cite{Welp1992,Takeshita2004,Ni2008,Guinea2009,Fisher2011,Chu2010,Tanatar2009,Tanatar2010,Blomberg2012,Hicks2014b} Uniaxial pressure has been used widely to detwin single crystal samples. For example, cooling iron pnictides superconductors through the tetragonal-orthorhombic structural transition with applied uniaxial pressure removes one of the twin domains, allowing anisotropic transport properties in the FeAs plane to be studied.\cite{Fisher2011,Chu2010,Tanatar2009,Tanatar2010,Blomberg2012} Uniaxial pressure can also directly modify electronic properties in graphene \cite{Ni2008,Guinea2009} as well as the superconducting transition temperature in $\rm Sr_2RuO_4$.\cite{Hicks2014b} Detwinned cuprate single crystal samples were obtained in a similar manner for the investigation of anisotropic electronic properties in the CuO$_2$ planes.\cite{Welp1992,Lavrov2001,Lavrov2002,Ando2002,Takeshita2004} Dramatic shifts of the superconducting transition temperature were also observed in $\rm La_{1.64}Eu_{0.2}Sr_{0.16}CuO_4$ \cite{Takeshita2004} and $\rm YBa_2Cu_3O_{7-\delta}$.\cite{Welp1992}

The focus of these earlier studies was mostly on the production of the sample to be used in low temperature transport studies, and quantitative aspects of the structural response to the application of the uniaxial pressure were not discussed in detail.\cite{Dhital2012} In this Letter, we examine how the twin structure in $\rm La_{2-x}Sr_xCuO_4$ (LSCO) responds to applied uniaxial pressure by using x-ray microdiffraction. We find that the crystal can be dynamically detwinned by applying only moderate compressive uniaxial pressure even at room temperature, which is well below the structural phase transition temperature. The twin domain structure recovers when the applied uniaxial pressure is removed in this case. What is surprising is that the compressive uniaxial pressure eliminates the twin domain with the {\em shorter} lattice parameter along the uniaxial pressure direction. That is, the lattice parameter becomes longer under compressive strain in the orthorhombic phase of LSCO. On the other hand, the effect of tensile strain on the twin domain structure is rather weak. Our findings illustrate that the response to uniaxial pressure is quite complicated and can quickly deviate from the linear-response regime even with moderate pressure.

%%%%%%%%%%%%%%%%%%%%%%%%%%
\begin{figure}[!b]
		\includegraphics[scale=0.7]{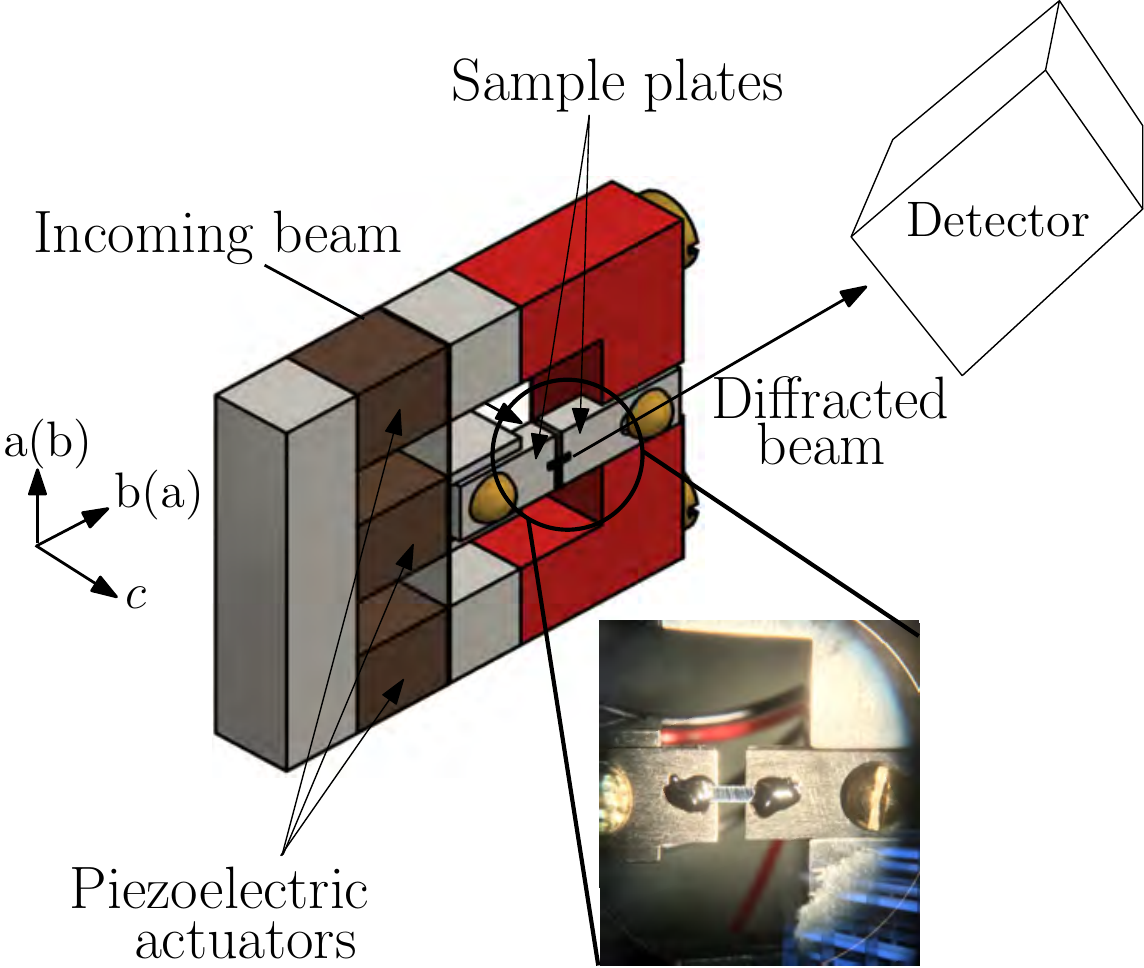}
				\caption{\label{fig:geo} Transmission geometry and orientation of the sample. The beam enters the gap between the sample plates from the back of the device and exists through the front. The sample thickness is approximately 0.1 mm, and the exposed surface available for diffraction is approximately 0.5 mm $\times$ 0.5 mm. \textit{Inset}: an example of the sample mounted between the sample plates.}
\end{figure}
%%%%%%%%%%%%%%%%%%%%%%%%%%%%%				

%%%%%%%%%%%%%%%%%%%%%%%%%%%%%				
\begin{figure}
		\includegraphics[scale=0.64]{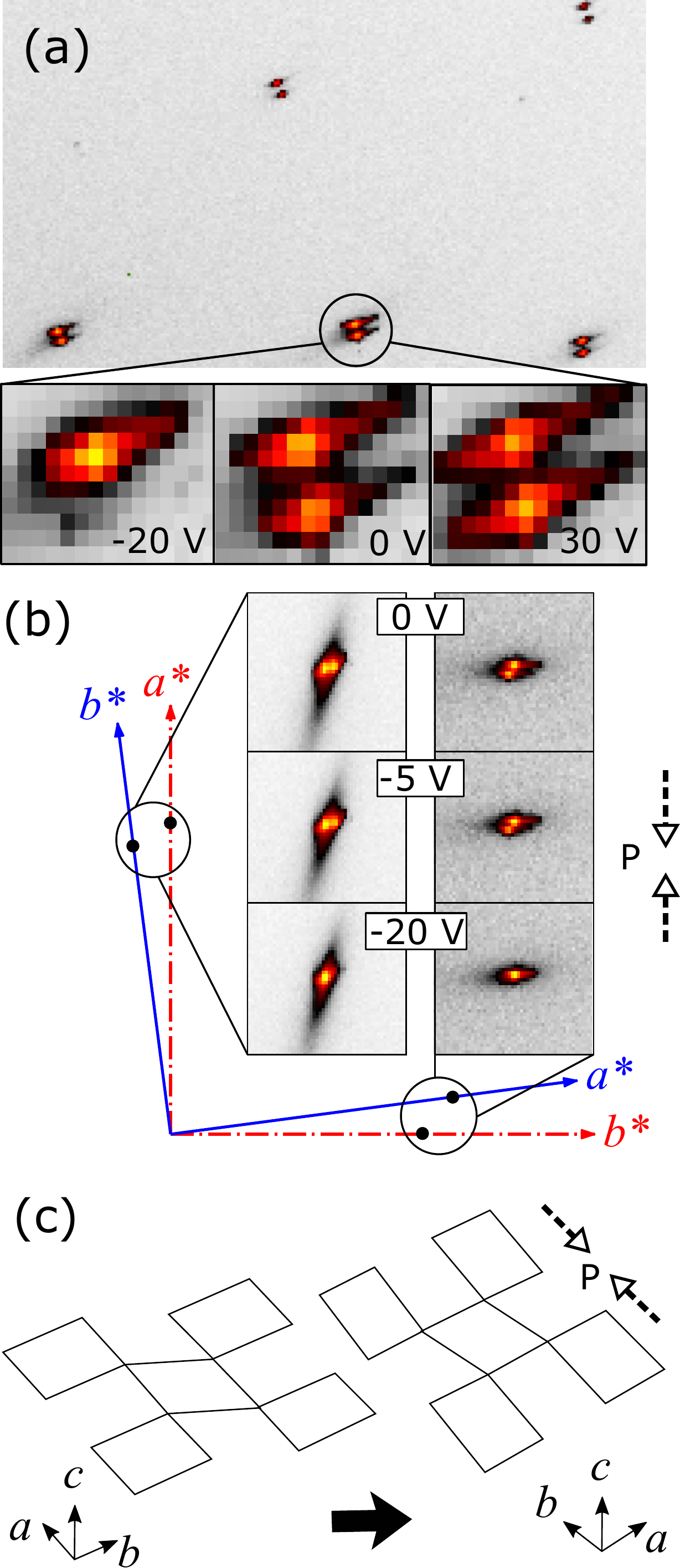}
				\caption{\label{fig:diff} (a) Partial view of the full diffraction pattern under no strain, showing five pairs of diffraction peaks. Also shown are the close-up views at -20 V (compressive strain), 0 V (no strain), and +30 V (tensile strain). (b) The reciprocal space perpendicular to the c-axis for the two twin domains (different colored axes). The filled circles are the Bragg peak positions for (200) and (020) peaks for the two domains. The evolution of the peak intensity with compressive strain is shown in the middle. (c) Schematic real-space picture illustrating the response of CuO$_2$ planes in LSCO under compressive strain. The buckled rectagles represent CuO$_6$ octahedra. Note the elongation along the compression direction.}
\end{figure}
%%%%%%%%%%%%%%%%%%%%%%%%%%%%%				

A superconducting $\rm La_{1.93}Sr_{0.07}CuO_4$ (LSCO x=0.07) crystal was used in our study, which possesses a layered perovskite structure with an orthorhombic to tetragonal structural phase transition at a temperature $T_s \approx 400$~K.\cite{Chen1991} In the high temperature regime, the LSCO sample is in tetragonal structure with the $I4/mmm$ symmetry. Below $T_s$, the crystal symmetry becomes $Bmab$, and the structure becomes heavily twinned. Earlier optical and electron microscopy studies reported stripe-like twin domains running along the orthorhombic \{110\} direction; the size (width) of these twin domains ranged from a few hundred nanometers to tens of microns.\cite{Sawada1989,Chen1991,Chen1993,Roy1991,Lavrov2002} It was also reported that there existed micro-twins within these bigger twin domains. That is, smaller micro domains of the other twin structure exists within one twin domain.\cite{Chen1991} Neutron and x-ray diffraction can also provide important quantitative information even without the spatial resolution of these microscopy techniques because of their bulk-sensitivity.\cite{Braden1992} The twinning gives rise to the splitting of a structural Bragg peak, with the relative intensity of each split peak representing the volume fraction of each twin domain. In this study, we use x-ray microdiffraction in Laue geometry to ensure that we are sensitive to bulk properties of the sample studied while maintaining the spatial resolution.

The x-ray diffraction experiments were carried out at the VESPERS beamline of the Canadian Light Source (CLS).
VESPERS is a bending magnet beamline with Kirkpatrick-Baez (KB) mirror-optics, with an effective beam spot of approximately
(3$\times$3) $\mu$m$^2$ over a wide energy range. We used an effectively ``white" beam with only a harmonic rejection mirror and no monochromator. The beam divergence was about 3 mrad. The Pilatus 1M detector (Dectris) was placed behind the sample to collect data in Laue geometry as illustrated in Fig.~\ref{fig:geo}. The sample-detector distance was calibrated with a thin silicon crystal and found to be 150~mm. The design of the uniaxial strain device used in our experiment was adapted from Hicks et al.\cite{Hicks2014} It utilizes three piezoelectric actuator stacks (PSt 150/7x7/7 from Piezomechanik GmbH), each 9 mm long and capable of a maximum extension of 9 $\mu$m at 150 V applied voltage. The three-leg design allows both compressive and tensile stress to be applied to the sample mounted across the two metal plates as shown in the inset photo in Fig.~\ref{fig:geo}.
All samples in this experiment were mounted using Stycast2850FT epoxy with 23LV catalyst, cured at 65 $^\circ C$ for two hours. Prior to each mounting, the sample plate surfaces were abraded with emery cloth to remove possible oxide layer formations and wiped off with acetone. The LSCO sample was cleaved to have the $c$ axis perpendicular to the sample plates and the $b(a)$ axis along the straining direction. The sample thickness was approximately 100~$\mu$m.

In principle, x-ray diffraction can be directly used to measure the strain on the sample. However, a very high angular resolution is necessary to detect the level of strain studied in our experiment: $10^{-4} \sim 10^{-3}$. We used two methods to estimate the sample strain. One is an indirect estimation given by $\epsilon_{P} (L_{P}/L_{S})$, where $\epsilon_{P}$ is the intrinsic strain of the piezo stack, and $L_{P}$ and $L_{S}$ are the lengths of the piezo stack (9~mm) and the sample respectively. Typical sample size was about 1~mm, which means that an order of magnitude larger strain can be achieved on the sample than that can be delivered by a piezo stack.
The second method utilized a silicon reference sample in the same diffraction setup. Although the angular resolution was not very high, the Laue geometry and a perfect crystal sample allow us to refine sample strain (see Supplementary Material). The sample strain along the uniaxial pressure direction was determined to be about $3 \times 10^{-4}$ when +25 V was applied to the piezo stack. Since the bulk modulus of silicon (100 GPa) is not too different from that of LSCO (130~GPa),\cite{Takahashi1994} one can then estimate the uniaxial pressure being applied in our LSCO experiment to be about 1 MPa/V. However, due to the large uncertainty in this estimate, we quote the voltage applied to the piezo stack in this paper without converting to pressure. A negative voltage $v$ represents a voltage $|v|$ applied to the middle piezo (compressive strain), while a positive voltage represents a voltage applied to the outer piezos (tensile strain).  Throughout this paper, we use the convention that a $(+)$ sign means tensile strain and a $(-)$ sign means compressive strain.

In Fig.~\ref{fig:diff} (a), we show a portion of the diffraction pattern of LSCO taken at the center of the sample under no stress. Five pairs of diffraction peaks due to twin domains can be seen. (Full diffraction images can be found in Supplementary Material.) We also show close-up images of a select pair of peaks under various strains: $-20$ V, 0 V, and +30 V respectively.
Before application of pressure (0 V), both domains are present and a pair of peaks are observed. At -20 V, the bottom peak vanishes and the sample becomes single domain at this location. At +30 V, the bottom peak intensifies slightly and there seems to be some change in the domain population.

The Laue scattering geometry in which the incoming x-ray beam is parallel to the c-axis is advantageous for finding out which domain each peak belongs. We show the schematic reciprocal space projection on the detector plane in Fig.~\ref{fig:diff}(b). Here the reciprocal space coordinate corresponding to the $a(b)$-axis oriented horizontally is shown in a blue solid line (red dot-dash line). The rotation angle between the two coordinate systems is exaggerated in this figure. The pair of peaks along these axes correspond to the (200)/(020) pair coming from the two twin domains. The diffraction data obtained at 0~V, -5~V, and -20~V for each pair are shown in the center panels, allowing one to follow the peaks and see which of them disappears when compressed. Between the two peaks on the vertical axes, one can clearly see that the peak on the right disappears, while the bottom peak disappears on the horizontal axes. Both disappearing peaks belong to the same twin domain (red dot-dash), as expected. But what is surprising is the direction of applied compressive pressure (dashed arrows). Our results indicate that the compressive strain stabilizes the twin domain with the {\em longer} lattice parameter (i.e., $b$) along the compression direction. That is, the twinned crystal of LSCO displays a negative compressibility. This will be discussed further below.

\begin{figure}
		\includegraphics[scale=0.53]{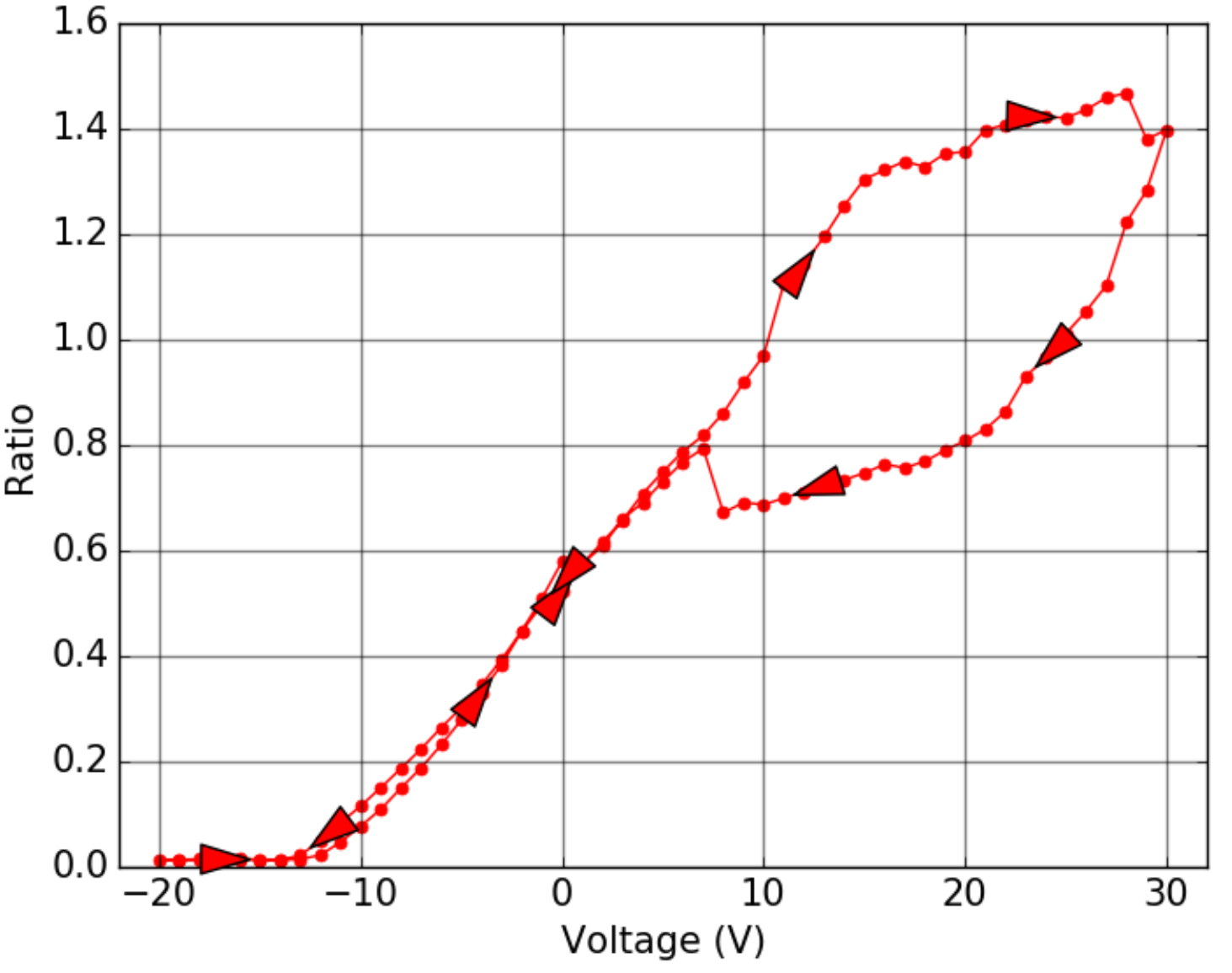}
				\caption{\label{fig:hys} The domain population ratio at the location where Fig.~\ref{fig:diff} (a) was taken, plotted as a function of voltage. The voltages applied were, in order, from 0 V to 30 V, from 30 V down to -20 V, and finally from -20 V back to 0 V, creating a full hysteresis loop. Here -10 V roughly corresponds to 10 MPa of compressive pressure (See text.)}
				\end{figure}

The domain populations appeared to respond to compressive and tensile strain very differently. To characterize this difference, the pressure (voltage) dependence of peak intensity ratio was studied. For a given voltage, the ratio of the integrated intensity of the bottom peak to the top peak was calculated for each of the five pairs of diffraction peaks in Fig.~\ref{fig:diff}(a). This number was then averaged over the five pairs and plotted against the voltage as shown in Fig.~\ref{fig:hys}. Now the asymmetry is made apparent. A $-20$ V compressive strain easily eliminates one domain, while both domains persist up to 30 V tensile strain. In addition, substantial hysteresis is observed in the case of tensile strain. We note that the piezo stack itself shows a mild hysterectic behavior, but in the opposite direction. Therefore, we believe that the observed hysteresis behavior is a characteristic of the sample under tensile strain.

						\begin{figure*}[t]
		\includegraphics[scale=0.8]{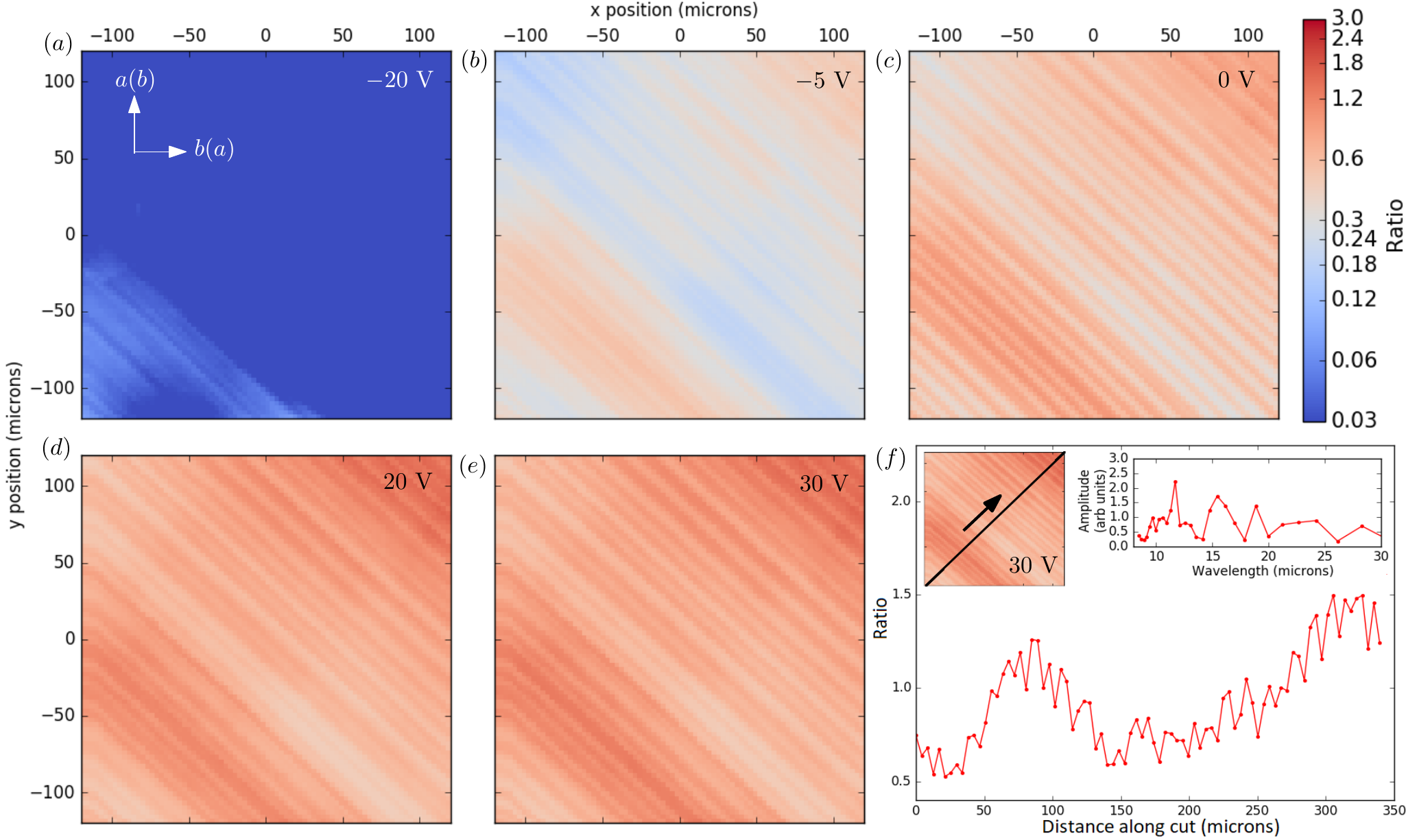}
				\caption{\label{fig:data} Domain population ratio map at various voltages. \textbf{(a)-(e)} Domain population ratio map for $-20$ V, $-5$ V, 0 V, 20 V, and 30 V. The scale of the color bar is logarithmic. \textbf{(f)} A plot of the ratio along a cut perpendicular to the striping pattern in the 30 V map. \textit{Inset:} An indication of the location of the cut and the Fourier transform of the cut in the wavelength domain.}
				\end{figure*}

In order to study spatial distribution of twin domains, we scanned the sample in the $ab$-plane over a 240 $\mu$m$\times$240$\mu$m area near the center of the sample. The step size used roughly matches the beam size (3 $\mu$m $\times$ 3 $\mu$m). At each point the same diffraction pattern as shown in Fig.~\ref{fig:diff} was obtained, and the averaged ratio of the 5 pairs was obtained in the same manner as before. A pseudocolor map presented in Fig.~\ref{fig:data} was then generated using these averaged ratios as intensities, giving a characterization of the domain population distribution in the sample. The stripy pattern in Fig.~\ref{fig:data} arises from the twin domains. We observe that the domain walls run along the $(1\bar{1}0)$ direction, and their separation is on the order of 10 microns. This is consistent with earlier observations from optical or electron microscopy \cite{Sawada1989,Chen1991,Chen1993,Roy1991,Lavrov2002}. The twin domain structure observed after the removal of uniaxial pressure seems to be identical to the original domain structure before the application of uniaxial pressure.

In Fig.~\ref{fig:data} (f), we plot the ratio along a line perpendicular to the domain walls in the 30 V map and its Fourier transform. The absence of any particular periodicity confirms the random nature of the domain size, which are presumably pinned by lattice defects. If the domains are much smaller than the spatial resolution determined by the beam size, then we expect a more homogeneous intensity distribution due to spatial averaging. The fact that we see these variations suggest that the domain size is at least comparable to the beam size. In addition, the presence of these patterns also mean that the domains are fairly well correlated along the c-direction. Significant variation in the domain population along the $c$ direction would result in a featureless map in this transmission geometry as an average is taken when the beam traverses through the sample of 100 $\mu$m thickness. We also note that the intensity ratio variation (e.g. for 0 V data) is much smaller than what is expected from an ideal domain distribution (zero to infinity in this ideal case). This could be due to the microtwins found inside larger twin domains as reported in Ref.~\cite{Chen1991}. To ensure that the observed pattern is an intrinsic sample property, we repeated the measurements with the reverse experimental geometry. When the strain device was rotated by 180 degrees about the vertical axis (the a(b) axis in Fig.~\ref{fig:geo}) so that the sample front is facing the beam, the observed domain pattern also flips as expected, confirming the intrinsic nature of the observed pattern. (Supplementary Material.)

To summarize, we show that application of compressive uniaxial pressure along the orthorhombic $a$ or $b$ axis of $\rm La_{1.93}Sr_{0.07}CuO_4$ can completely eliminate the twin domain with the $a$-direction along the uniaxial direction. This is a very surprising discovery by virtue of the fact that $a$ is smaller than $b$. So why does the crystal prefer to elongate under uniaxial compression? Although a quantitative theory is lacking, we can gain some qualitative insights by examining why $\rm La_2CuO_4$ becomes orthorhombic below 450~K in the first place. This is often attributed to bond-length mismatch \cite{Goodenough1991}. The Cu-O equilibrium bond length is too large compared to the equilibrium bond length of La-O, placing the CuO$_2$ layer under compressive stress and the LaO layer under tension even in the absence of applied pressure. Buckling of the CuO$_6$ octahedra addresses this problem because tilting of a CuO$_6$ octahedron brings the apical oxygen closer to La without shortening the Cu-O bond length. Therefore, apparently, application of compressive stress to (already stressed) Cu-O bonds along the $a$-axis causes them to buckle and become the $b$-axis (See Fig. 2(c) for a schematic illustration). The buckling also distorts the CuO$_6$ octahedra, which causes the $b$-axis to have a larger lattice parameter. In fact, rotation of rigid structural units plays an important role in many materials exhibiting negative compressibility.\cite{Baughman1998,Attard2016}
It should be noted that the layered structure of $\rm La_{2-x}Sr_{x}CuO_4$ makes this a particularly simple system to study physics of octahedral tilting and uniaxial strain, unlike three-dimensional structures such as perovskites, in which octahedral tilting and rotation in response to pressure can be quite complicated.

Our findings have several interesting implications for future studies. We demonstrated the power of x-ray Laue microdiffraction as a tool for studying strain in materials. In particular, the combination of spatial resolution and bulk-sensitivity allows one to study the response of twin domains in detail. Further investigation of various negative compressibility materials could be useful for elucidating non-linear response of materials under uniaxial pressure.

\section*{Supplementary Material}
See Supplementary Material for the strain refinement of the silicon reference sample, the reverse experimental geometry data, and the full diffraction image of the peaks shown in Fig. 2(a).

\begin{acknowledgments}
Work at the University of Toronto was supported by the Natural Science and Engineering Research Council (NSERC) of Canada through Discovery Grant (RGPIN-2014-06071), Canadian Foundation for Innovation, and Ontario Innovation Trust. Research performed at the Canadian Light Source is supported by the Canada Foundation for Innovation, NSERC, the University of Saskatchewan, the Government of Saskatchewan, Western Economic Diversification Canada, the National Research Council Canada, and the Canadian Institute of Health Research. X. Y. Z. acknowledges the receipt of support from the NSERC Undergraduate Student Research Award.% Put your acknowledgments here.
\end{acknowledgments}

%\bibliography{strain}

\begin{thebibliography}{25}%
\makeatletter
\providecommand \@ifxundefined [1]{%
 \@ifx{#1\undefined}
}%
\providecommand \@ifnum [1]{%
 \ifnum #1\expandafter \@firstoftwo
 \else \expandafter \@secondoftwo
 \fi
}%
\providecommand \@ifx [1]{%
 \ifx #1\expandafter \@firstoftwo
 \else \expandafter \@secondoftwo
 \fi
}%
\providecommand \natexlab [1]{#1}%
\providecommand \enquote  [1]{``#1''}%
\providecommand \bibnamefont  [1]{#1}%
\providecommand \bibfnamefont [1]{#1}%
\providecommand \citenamefont [1]{#1}%
\providecommand \href@noop [0]{\@secondoftwo}%
\providecommand \href [0]{\begingroup \@sanitize@url \@href}%
\providecommand \@href[1]{\@@startlink{#1}\@@href}%
\providecommand \@@href[1]{\endgroup#1\@@endlink}%
\providecommand \@sanitize@url [0]{\catcode `\\12\catcode `\$12\catcode
  `\&12\catcode `\#12\catcode `\^12\catcode `\_12\catcode `\%12\relax}%
\providecommand \@@startlink[1]{}%
\providecommand \@@endlink[0]{}%
\providecommand \url  [0]{\begingroup\@sanitize@url \@url }%
\providecommand \@url [1]{\endgroup\@href {#1}{\urlprefix }}%
\providecommand \urlprefix  [0]{URL }%
\providecommand \Eprint [0]{\href }%
\providecommand \doibase [0]{http://dx.doi.org/}%
\providecommand \selectlanguage [0]{\@gobble}%
\providecommand \bibinfo  [0]{\@secondoftwo}%
\providecommand \bibfield  [0]{\@secondoftwo}%
\providecommand \translation [1]{[#1]}%
\providecommand \BibitemOpen [0]{}%
\providecommand \bibitemStop [0]{}%
\providecommand \bibitemNoStop [0]{.\EOS\space}%
\providecommand \EOS [0]{\spacefactor3000\relax}%
\providecommand \BibitemShut  [1]{\csname bibitem#1\endcsname}%
\let\auto@bib@innerbib\@empty
%</preamble>
\bibitem [{Note1()}]{Note1}%
  \BibitemOpen
  \bibinfo {note} {We use the term pressure in our experiment, since both
  stress and strain are second-rank tensor quantities, and thus not easy to
  define uniaxial direction.}\BibitemShut {Stop}%
\bibitem [{\citenamefont {Welp}\ \emph {et~al.}(1992)\citenamefont {Welp},
  \citenamefont {Grimsditch}, \citenamefont {Fleshler}, \citenamefont
  {Nessler}, \citenamefont {Downey}, \citenamefont {Crabtree},\ and\
  \citenamefont {Guimpel}}]{Welp1992}%
  \BibitemOpen
  \bibfield  {author} {\bibinfo {author} {\bibfnamefont {U.}~\bibnamefont
  {Welp}}, \bibinfo {author} {\bibfnamefont {M.}~\bibnamefont {Grimsditch}},
  \bibinfo {author} {\bibfnamefont {S.}~\bibnamefont {Fleshler}}, \bibinfo
  {author} {\bibfnamefont {W.}~\bibnamefont {Nessler}}, \bibinfo {author}
  {\bibfnamefont {J.}~\bibnamefont {Downey}}, \bibinfo {author} {\bibfnamefont
  {G.~W.}\ \bibnamefont {Crabtree}}, \ and\ \bibinfo {author} {\bibfnamefont
  {J.}~\bibnamefont {Guimpel}},\ }\href {\doibase 10.1103/PhysRevLett.69.2130}
  {\bibfield  {journal} {\bibinfo  {journal} {Phys. Rev. Lett.}\ }\textbf
  {\bibinfo {volume} {69}},\ \bibinfo {pages} {2130} (\bibinfo {year}
  {1992})}\BibitemShut {NoStop}%
\bibitem [{\citenamefont {Takeshita}\ \emph {et~al.}(2004)\citenamefont
  {Takeshita}, \citenamefont {Sasagawa}, \citenamefont {Sugioka}, \citenamefont
  {Tokura},\ and\ \citenamefont {Takagi}}]{Takeshita2004}%
  \BibitemOpen
  \bibfield  {author} {\bibinfo {author} {\bibfnamefont {N.}~\bibnamefont
  {Takeshita}}, \bibinfo {author} {\bibfnamefont {T.}~\bibnamefont {Sasagawa}},
  \bibinfo {author} {\bibfnamefont {T.}~\bibnamefont {Sugioka}}, \bibinfo
  {author} {\bibfnamefont {Y.}~\bibnamefont {Tokura}}, \ and\ \bibinfo {author}
  {\bibfnamefont {H.}~\bibnamefont {Takagi}},\ }\href {\doibase
  10.1143/JPSJ.73.1123} {\bibfield  {journal} {\bibinfo  {journal} {Journal of
  the Physical Society of Japan}\ }\textbf {\bibinfo {volume} {73}},\ \bibinfo
  {pages} {1123} (\bibinfo {year} {2004})},\ \Eprint
  {http://arxiv.org/abs/https://doi.org/10.1143/JPSJ.73.1123}
  {https://doi.org/10.1143/JPSJ.73.1123} \BibitemShut {NoStop}%
\bibitem [{\citenamefont {Ni}\ \emph {et~al.}(2008)\citenamefont {Ni},
  \citenamefont {Yu}, \citenamefont {Lu}, \citenamefont {Wang}, \citenamefont
  {Feng},\ and\ \citenamefont {Shen}}]{Ni2008}%
  \BibitemOpen
  \bibfield  {author} {\bibinfo {author} {\bibfnamefont {Z.~H.}\ \bibnamefont
  {Ni}}, \bibinfo {author} {\bibfnamefont {T.}~\bibnamefont {Yu}}, \bibinfo
  {author} {\bibfnamefont {Y.~H.}\ \bibnamefont {Lu}}, \bibinfo {author}
  {\bibfnamefont {Y.~Y.}\ \bibnamefont {Wang}}, \bibinfo {author}
  {\bibfnamefont {Y.~P.}\ \bibnamefont {Feng}}, \ and\ \bibinfo {author}
  {\bibfnamefont {Z.~X.}\ \bibnamefont {Shen}},\ }\href {\doibase
  10.1021/nn800459e} {\bibfield  {journal} {\bibinfo  {journal} {ACS Nano}\
  }\textbf {\bibinfo {volume} {2}},\ \bibinfo {pages} {2301} (\bibinfo {year}
  {2008})},\ \bibinfo {note} {pMID: 19206396},\ \Eprint
  {http://arxiv.org/abs/https://doi.org/10.1021/nn800459e}
  {https://doi.org/10.1021/nn800459e} \BibitemShut {NoStop}%
\bibitem [{\citenamefont {Guinea}, \citenamefont {Katsnelson},\ and\
  \citenamefont {Geim}(2009)}]{Guinea2009}%
  \BibitemOpen
  \bibfield  {author} {\bibinfo {author} {\bibfnamefont {F.}~\bibnamefont
  {Guinea}}, \bibinfo {author} {\bibfnamefont {M.~I.}\ \bibnamefont
  {Katsnelson}}, \ and\ \bibinfo {author} {\bibfnamefont {A.~K.}\ \bibnamefont
  {Geim}},\ }\href {http://dx.doi.org/10.1038/nphys1420} {\bibfield  {journal}
  {\bibinfo  {journal} {Nature Physics}\ }\textbf {\bibinfo {volume} {6}},\
  \bibinfo {pages} {30} (\bibinfo {year} {2009})}\BibitemShut {NoStop}%
\bibitem [{\citenamefont {Fisher}, \citenamefont {Degiorgi},\ and\
  \citenamefont {Shen}(2011)}]{Fisher2011}%
  \BibitemOpen
  \bibfield  {author} {\bibinfo {author} {\bibfnamefont {I.~R.}\ \bibnamefont
  {Fisher}}, \bibinfo {author} {\bibfnamefont {L.}~\bibnamefont {Degiorgi}}, \
  and\ \bibinfo {author} {\bibfnamefont {Z.~X.}\ \bibnamefont {Shen}},\ }\href
  {http://stacks.iop.org/0034-4885/74/i=12/a=124506} {\bibfield  {journal}
  {\bibinfo  {journal} {Reports on Progress in Physics}\ }\textbf {\bibinfo
  {volume} {74}},\ \bibinfo {pages} {124506} (\bibinfo {year}
  {2011})}\BibitemShut {NoStop}%
\bibitem [{\citenamefont {Chu}\ \emph {et~al.}(2010)\citenamefont {Chu},
  \citenamefont {Analytis}, \citenamefont {De~Greve}, \citenamefont {McMahon},
  \citenamefont {Islam}, \citenamefont {Yamamoto},\ and\ \citenamefont
  {Fisher}}]{Chu2010}%
  \BibitemOpen
  \bibfield  {author} {\bibinfo {author} {\bibfnamefont {J.-H.}\ \bibnamefont
  {Chu}}, \bibinfo {author} {\bibfnamefont {J.~G.}\ \bibnamefont {Analytis}},
  \bibinfo {author} {\bibfnamefont {K.}~\bibnamefont {De~Greve}}, \bibinfo
  {author} {\bibfnamefont {P.~L.}\ \bibnamefont {McMahon}}, \bibinfo {author}
  {\bibfnamefont {Z.}~\bibnamefont {Islam}}, \bibinfo {author} {\bibfnamefont
  {Y.}~\bibnamefont {Yamamoto}}, \ and\ \bibinfo {author} {\bibfnamefont
  {I.~R.}\ \bibnamefont {Fisher}},\ }\href {\doibase 10.1126/science.1190482}
  {\bibfield  {journal} {\bibinfo  {journal} {Science}\ }\textbf {\bibinfo
  {volume} {329}},\ \bibinfo {pages} {824} (\bibinfo {year} {2010})},\ \Eprint
  {http://arxiv.org/abs/http://science.sciencemag.org/content/329/5993/824.full.pdf}
  {http://science.sciencemag.org/content/329/5993/824.full.pdf} \BibitemShut
  {NoStop}%
\bibitem [{\citenamefont {Tanatar}\ \emph {et~al.}(2009)\citenamefont
  {Tanatar}, \citenamefont {Kreyssig}, \citenamefont {Nandi}, \citenamefont
  {Ni}, \citenamefont {Bud'ko}, \citenamefont {Canfield}, \citenamefont
  {Goldman},\ and\ \citenamefont {Prozorov}}]{Tanatar2009}%
  \BibitemOpen
  \bibfield  {author} {\bibinfo {author} {\bibfnamefont {M.~A.}\ \bibnamefont
  {Tanatar}}, \bibinfo {author} {\bibfnamefont {A.}~\bibnamefont {Kreyssig}},
  \bibinfo {author} {\bibfnamefont {S.}~\bibnamefont {Nandi}}, \bibinfo
  {author} {\bibfnamefont {N.}~\bibnamefont {Ni}}, \bibinfo {author}
  {\bibfnamefont {S.~L.}\ \bibnamefont {Bud'ko}}, \bibinfo {author}
  {\bibfnamefont {P.~C.}\ \bibnamefont {Canfield}}, \bibinfo {author}
  {\bibfnamefont {A.~I.}\ \bibnamefont {Goldman}}, \ and\ \bibinfo {author}
  {\bibfnamefont {R.}~\bibnamefont {Prozorov}},\ }\href {\doibase
  10.1103/PhysRevB.79.180508} {\bibfield  {journal} {\bibinfo  {journal} {Phys.
  Rev. B}\ }\textbf {\bibinfo {volume} {79}},\ \bibinfo {pages} {180508}
  (\bibinfo {year} {2009})}\BibitemShut {NoStop}%
\bibitem [{\citenamefont {Tanatar}\ \emph {et~al.}(2010)\citenamefont
  {Tanatar}, \citenamefont {Blomberg}, \citenamefont {Kreyssig}, \citenamefont
  {Kim}, \citenamefont {Ni}, \citenamefont {Thaler}, \citenamefont {Bud'ko},
  \citenamefont {Canfield}, \citenamefont {Goldman}, \citenamefont {Mazin},\
  and\ \citenamefont {Prozorov}}]{Tanatar2010}%
  \BibitemOpen
  \bibfield  {author} {\bibinfo {author} {\bibfnamefont {M.~A.}\ \bibnamefont
  {Tanatar}}, \bibinfo {author} {\bibfnamefont {E.~C.}\ \bibnamefont
  {Blomberg}}, \bibinfo {author} {\bibfnamefont {A.}~\bibnamefont {Kreyssig}},
  \bibinfo {author} {\bibfnamefont {M.~G.}\ \bibnamefont {Kim}}, \bibinfo
  {author} {\bibfnamefont {N.}~\bibnamefont {Ni}}, \bibinfo {author}
  {\bibfnamefont {A.}~\bibnamefont {Thaler}}, \bibinfo {author} {\bibfnamefont
  {S.~L.}\ \bibnamefont {Bud'ko}}, \bibinfo {author} {\bibfnamefont {P.~C.}\
  \bibnamefont {Canfield}}, \bibinfo {author} {\bibfnamefont {A.~I.}\
  \bibnamefont {Goldman}}, \bibinfo {author} {\bibfnamefont {I.~I.}\
  \bibnamefont {Mazin}}, \ and\ \bibinfo {author} {\bibfnamefont
  {R.}~\bibnamefont {Prozorov}},\ }\href {\doibase 10.1103/PhysRevB.81.184508}
  {\bibfield  {journal} {\bibinfo  {journal} {Phys. Rev. B}\ }\textbf {\bibinfo
  {volume} {81}},\ \bibinfo {pages} {184508} (\bibinfo {year}
  {2010})}\BibitemShut {NoStop}%
\bibitem [{\citenamefont {Blomberg}\ \emph {et~al.}(2012)\citenamefont
  {Blomberg}, \citenamefont {Kreyssig}, \citenamefont {Tanatar}, \citenamefont
  {Fernandes}, \citenamefont {Kim}, \citenamefont {Thaler}, \citenamefont
  {Schmalian}, \citenamefont {Bud'ko}, \citenamefont {Canfield}, \citenamefont
  {Goldman},\ and\ \citenamefont {Prozorov}}]{Blomberg2012}%
  \BibitemOpen
  \bibfield  {author} {\bibinfo {author} {\bibfnamefont {E.~C.}\ \bibnamefont
  {Blomberg}}, \bibinfo {author} {\bibfnamefont {A.}~\bibnamefont {Kreyssig}},
  \bibinfo {author} {\bibfnamefont {M.~A.}\ \bibnamefont {Tanatar}}, \bibinfo
  {author} {\bibfnamefont {R.~M.}\ \bibnamefont {Fernandes}}, \bibinfo {author}
  {\bibfnamefont {M.~G.}\ \bibnamefont {Kim}}, \bibinfo {author} {\bibfnamefont
  {A.}~\bibnamefont {Thaler}}, \bibinfo {author} {\bibfnamefont
  {J.}~\bibnamefont {Schmalian}}, \bibinfo {author} {\bibfnamefont {S.~L.}\
  \bibnamefont {Bud'ko}}, \bibinfo {author} {\bibfnamefont {P.~C.}\
  \bibnamefont {Canfield}}, \bibinfo {author} {\bibfnamefont {A.~I.}\
  \bibnamefont {Goldman}}, \ and\ \bibinfo {author} {\bibfnamefont
  {R.}~\bibnamefont {Prozorov}},\ }\href {\doibase 10.1103/PhysRevB.85.144509}
  {\bibfield  {journal} {\bibinfo  {journal} {Phys. Rev. B}\ }\textbf {\bibinfo
  {volume} {85}},\ \bibinfo {pages} {144509} (\bibinfo {year}
  {2012})}\BibitemShut {NoStop}%
\bibitem [{\citenamefont {Hicks}\ \emph
  {et~al.}(2014{\natexlab{a}})\citenamefont {Hicks}, \citenamefont {Brodsky},
  \citenamefont {Yelland}, \citenamefont {Gibbs}, \citenamefont {Bruin},
  \citenamefont {Barber}, \citenamefont {Edkins}, \citenamefont {Nishimura},
  \citenamefont {Yonezawa}, \citenamefont {Maeno},\ and\ \citenamefont
  {Mackenzie}}]{Hicks2014b}%
  \BibitemOpen
  \bibfield  {author} {\bibinfo {author} {\bibfnamefont {C.~W.}\ \bibnamefont
  {Hicks}}, \bibinfo {author} {\bibfnamefont {D.~O.}\ \bibnamefont {Brodsky}},
  \bibinfo {author} {\bibfnamefont {E.~A.}\ \bibnamefont {Yelland}}, \bibinfo
  {author} {\bibfnamefont {A.~S.}\ \bibnamefont {Gibbs}}, \bibinfo {author}
  {\bibfnamefont {J.~A.~N.}\ \bibnamefont {Bruin}}, \bibinfo {author}
  {\bibfnamefont {M.~E.}\ \bibnamefont {Barber}}, \bibinfo {author}
  {\bibfnamefont {S.~D.}\ \bibnamefont {Edkins}}, \bibinfo {author}
  {\bibfnamefont {K.}~\bibnamefont {Nishimura}}, \bibinfo {author}
  {\bibfnamefont {S.}~\bibnamefont {Yonezawa}}, \bibinfo {author}
  {\bibfnamefont {Y.}~\bibnamefont {Maeno}}, \ and\ \bibinfo {author}
  {\bibfnamefont {A.~P.}\ \bibnamefont {Mackenzie}},\ }\href {\doibase
  10.1126/science.1248292} {\bibfield  {journal} {\bibinfo  {journal}
  {Science}\ }\textbf {\bibinfo {volume} {344}},\ \bibinfo {pages} {283}
  (\bibinfo {year} {2014}{\natexlab{a}})},\ \Eprint
  {http://arxiv.org/abs/http://science.sciencemag.org/content/344/6181/283.full.pdf}
  {http://science.sciencemag.org/content/344/6181/283.full.pdf} \BibitemShut
  {NoStop}%
\bibitem [{\citenamefont {Lavrov}\ \emph {et~al.}(2001)\citenamefont {Lavrov},
  \citenamefont {Ando}, \citenamefont {Komiya},\ and\ \citenamefont
  {Tsukada}}]{Lavrov2001}%
  \BibitemOpen
  \bibfield  {author} {\bibinfo {author} {\bibfnamefont {A.~N.}\ \bibnamefont
  {Lavrov}}, \bibinfo {author} {\bibfnamefont {Y.}~\bibnamefont {Ando}},
  \bibinfo {author} {\bibfnamefont {S.}~\bibnamefont {Komiya}}, \ and\ \bibinfo
  {author} {\bibfnamefont {I.}~\bibnamefont {Tsukada}},\ }\href {\doibase
  10.1103/PhysRevLett.87.017007} {\bibfield  {journal} {\bibinfo  {journal}
  {Phys. Rev. Lett.}\ }\textbf {\bibinfo {volume} {87}},\ \bibinfo {pages}
  {017007} (\bibinfo {year} {2001})}\BibitemShut {NoStop}%
\bibitem [{\citenamefont {{Lavrov}}, \citenamefont {{Komiya}},\ and\
  \citenamefont {{Ando}}(2002)}]{Lavrov2002}%
  \BibitemOpen
  \bibfield  {author} {\bibinfo {author} {\bibfnamefont {A.~N.}\ \bibnamefont
  {{Lavrov}}}, \bibinfo {author} {\bibfnamefont {S.}~\bibnamefont {{Komiya}}},
  \ and\ \bibinfo {author} {\bibfnamefont {Y.}~\bibnamefont {{Ando}}},\ }\href
  {\doibase 10.1038/418385a} {\bibfield  {journal} {\bibinfo  {journal}
  {nature}\ }\textbf {\bibinfo {volume} {418}},\ \bibinfo {pages} {385}
  (\bibinfo {year} {2002})},\ \Eprint {http://arxiv.org/abs/cond-mat/0208013}
  {cond-mat/0208013} \BibitemShut {NoStop}%
\bibitem [{\citenamefont {Ando}\ \emph {et~al.}(2002)\citenamefont {Ando},
  \citenamefont {Segawa}, \citenamefont {Komiya},\ and\ \citenamefont
  {Lavrov}}]{Ando2002}%
  \BibitemOpen
  \bibfield  {author} {\bibinfo {author} {\bibfnamefont {Y.}~\bibnamefont
  {Ando}}, \bibinfo {author} {\bibfnamefont {K.}~\bibnamefont {Segawa}},
  \bibinfo {author} {\bibfnamefont {S.}~\bibnamefont {Komiya}}, \ and\ \bibinfo
  {author} {\bibfnamefont {A.~N.}\ \bibnamefont {Lavrov}},\ }\href {\doibase
  10.1103/PhysRevLett.88.137005} {\bibfield  {journal} {\bibinfo  {journal}
  {Phys. Rev. Lett.}\ }\textbf {\bibinfo {volume} {88}},\ \bibinfo {pages}
  {137005} (\bibinfo {year} {2002})}\BibitemShut {NoStop}%
\bibitem [{\citenamefont {Dhital}\ \emph {et~al.}(2012)\citenamefont {Dhital},
  \citenamefont {Yamani}, \citenamefont {Tian}, \citenamefont {Zeretsky},
  \citenamefont {Sefat}, \citenamefont {Wang}, \citenamefont {Birgeneau},\ and\
  \citenamefont {Wilson}}]{Dhital2012}%
  \BibitemOpen
  \bibfield  {author} {\bibinfo {author} {\bibfnamefont {C.}~\bibnamefont
  {Dhital}}, \bibinfo {author} {\bibfnamefont {Z.}~\bibnamefont {Yamani}},
  \bibinfo {author} {\bibfnamefont {W.}~\bibnamefont {Tian}}, \bibinfo {author}
  {\bibfnamefont {J.}~\bibnamefont {Zeretsky}}, \bibinfo {author}
  {\bibfnamefont {A.~S.}\ \bibnamefont {Sefat}}, \bibinfo {author}
  {\bibfnamefont {Z.}~\bibnamefont {Wang}}, \bibinfo {author} {\bibfnamefont
  {R.~J.}\ \bibnamefont {Birgeneau}}, \ and\ \bibinfo {author} {\bibfnamefont
  {S.~D.}\ \bibnamefont {Wilson}},\ }\href {\doibase
  10.1103/PhysRevLett.108.087001} {\bibfield  {journal} {\bibinfo  {journal}
  {Phys. Rev. Lett.}\ }\textbf {\bibinfo {volume} {108}},\ \bibinfo {pages}
  {087001} (\bibinfo {year} {2012})}\BibitemShut {NoStop}%
\bibitem [{\citenamefont {Chen}\ \emph {et~al.}(1991)\citenamefont {Chen},
  \citenamefont {Cheong}, \citenamefont {Werder}, \citenamefont {Cooper},\ and\
  \citenamefont {Rupp}}]{Chen1991}%
  \BibitemOpen
  \bibfield  {author} {\bibinfo {author} {\bibfnamefont {C.}~\bibnamefont
  {Chen}}, \bibinfo {author} {\bibfnamefont {S.-W.}\ \bibnamefont {Cheong}},
  \bibinfo {author} {\bibfnamefont {D.}~\bibnamefont {Werder}}, \bibinfo
  {author} {\bibfnamefont {A.}~\bibnamefont {Cooper}}, \ and\ \bibinfo {author}
  {\bibfnamefont {L.}~\bibnamefont {Rupp}},\ }\href {\doibase
  https://doi.org/10.1016/0921-4534(91)90601-T} {\bibfield  {journal} {\bibinfo
   {journal} {Physica C: Superconductivity}\ }\textbf {\bibinfo {volume}
  {175}},\ \bibinfo {pages} {301 } (\bibinfo {year} {1991})}\BibitemShut
  {NoStop}%
\bibitem [{\citenamefont {Sawada}\ \emph {et~al.}(1989)\citenamefont {Sawada},
  \citenamefont {Nishihata}, \citenamefont {Oka},\ and\ \citenamefont
  {Unoki}}]{Sawada1989}%
  \BibitemOpen
  \bibfield  {author} {\bibinfo {author} {\bibfnamefont {A.}~\bibnamefont
  {Sawada}}, \bibinfo {author} {\bibfnamefont {Y.}~\bibnamefont {Nishihata}},
  \bibinfo {author} {\bibfnamefont {K.}~\bibnamefont {Oka}}, \ and\ \bibinfo
  {author} {\bibfnamefont {H.}~\bibnamefont {Unoki}},\ }\href
  {http://stacks.iop.org/1347-4065/28/i=10A/a=L1787} {\bibfield  {journal}
  {\bibinfo  {journal} {Japanese Journal of Applied Physics}\ }\textbf
  {\bibinfo {volume} {28}},\ \bibinfo {pages} {L1787} (\bibinfo {year}
  {1989})}\BibitemShut {NoStop}%
\bibitem [{\citenamefont {Chen}(1993)}]{Chen1993}%
  \BibitemOpen
  \bibfield  {author} {\bibinfo {author} {\bibfnamefont {C.}~\bibnamefont
  {Chen}},\ }\href {\doibase doi:10.4028/www.scientific.net/MSF.137-139.257}
  {\bibfield  {journal} {\bibinfo  {journal} {Materials Science Forum}\
  }\textbf {\bibinfo {volume} {137-139}},\ \bibinfo {pages} {257} (\bibinfo
  {year} {1993})}\BibitemShut {NoStop}%
\bibitem [{\citenamefont {Roy}\ and\ \citenamefont {Mitchell}(1991)}]{Roy1991}%
  \BibitemOpen
  \bibfield  {author} {\bibinfo {author} {\bibfnamefont {T.}~\bibnamefont
  {Roy}}\ and\ \bibinfo {author} {\bibfnamefont {T.~E.}\ \bibnamefont
  {Mitchell}},\ }\href {\doibase 10.1080/01418619108204846} {\bibfield
  {journal} {\bibinfo  {journal} {Philosophical Magazine A}\ }\textbf {\bibinfo
  {volume} {63}},\ \bibinfo {pages} {225} (\bibinfo {year} {1991})},\ \Eprint
  {http://arxiv.org/abs/https://doi.org/10.1080/01418619108204846}
  {https://doi.org/10.1080/01418619108204846} \BibitemShut {NoStop}%
\bibitem [{\citenamefont {Braden}\ \emph {et~al.}(1992)\citenamefont {Braden},
  \citenamefont {Heger}, \citenamefont {Schweiss}, \citenamefont {Fisk},
  \citenamefont {Gamayunov}, \citenamefont {Tanaka},\ and\ \citenamefont
  {Kojima}}]{Braden1992}%
  \BibitemOpen
  \bibfield  {author} {\bibinfo {author} {\bibfnamefont {M.}~\bibnamefont
  {Braden}}, \bibinfo {author} {\bibfnamefont {G.}~\bibnamefont {Heger}},
  \bibinfo {author} {\bibfnamefont {P.}~\bibnamefont {Schweiss}}, \bibinfo
  {author} {\bibfnamefont {Z.}~\bibnamefont {Fisk}}, \bibinfo {author}
  {\bibfnamefont {K.}~\bibnamefont {Gamayunov}}, \bibinfo {author}
  {\bibfnamefont {I.}~\bibnamefont {Tanaka}}, \ and\ \bibinfo {author}
  {\bibfnamefont {H.}~\bibnamefont {Kojima}},\ }\href {\doibase
  https://doi.org/10.1016/0921-4534(92)90944-8} {\bibfield  {journal} {\bibinfo
   {journal} {Physica C: Superconductivity}\ }\textbf {\bibinfo {volume}
  {191}},\ \bibinfo {pages} {455 } (\bibinfo {year} {1992})}\BibitemShut
  {NoStop}%
\bibitem [{\citenamefont {Hicks}\ \emph
  {et~al.}(2014{\natexlab{b}})\citenamefont {Hicks}, \citenamefont {Barber},
  \citenamefont {Edkins}, \citenamefont {Brodsky},\ and\ \citenamefont
  {Mackenzie}}]{Hicks2014}%
  \BibitemOpen
  \bibfield  {author} {\bibinfo {author} {\bibfnamefont {C.~W.}\ \bibnamefont
  {Hicks}}, \bibinfo {author} {\bibfnamefont {M.~E.}\ \bibnamefont {Barber}},
  \bibinfo {author} {\bibfnamefont {S.~D.}\ \bibnamefont {Edkins}}, \bibinfo
  {author} {\bibfnamefont {D.~O.}\ \bibnamefont {Brodsky}}, \ and\ \bibinfo
  {author} {\bibfnamefont {A.~P.}\ \bibnamefont {Mackenzie}},\ }\href {\doibase
  10.1063/1.4881611} {\bibfield  {journal} {\bibinfo  {journal} {Review of
  Scientific Instruments}\ }\textbf {\bibinfo {volume} {85}},\ \bibinfo {pages}
  {065003} (\bibinfo {year} {2014}{\natexlab{b}})},\ \Eprint
  {http://arxiv.org/abs/https://doi.org/10.1063/1.4881611}
  {https://doi.org/10.1063/1.4881611} \BibitemShut {NoStop}%
\bibitem [{\citenamefont {Takahashi}\ \emph {et~al.}(1994)\citenamefont
  {Takahashi}, \citenamefont {Shaked}, \citenamefont {Hunter}, \citenamefont
  {Radaelli}, \citenamefont {Hitterman}, \citenamefont {Hinks},\ and\
  \citenamefont {Jorgensen}}]{Takahashi1994}%
  \BibitemOpen
  \bibfield  {author} {\bibinfo {author} {\bibfnamefont {H.}~\bibnamefont
  {Takahashi}}, \bibinfo {author} {\bibfnamefont {H.}~\bibnamefont {Shaked}},
  \bibinfo {author} {\bibfnamefont {B.~A.}\ \bibnamefont {Hunter}}, \bibinfo
  {author} {\bibfnamefont {P.~G.}\ \bibnamefont {Radaelli}}, \bibinfo {author}
  {\bibfnamefont {R.~L.}\ \bibnamefont {Hitterman}}, \bibinfo {author}
  {\bibfnamefont {D.~G.}\ \bibnamefont {Hinks}}, \ and\ \bibinfo {author}
  {\bibfnamefont {J.~D.}\ \bibnamefont {Jorgensen}},\ }\href {\doibase
  10.1103/PhysRevB.50.3221} {\bibfield  {journal} {\bibinfo  {journal} {Phys.
  Rev. B}\ }\textbf {\bibinfo {volume} {50}},\ \bibinfo {pages} {3221}
  (\bibinfo {year} {1994})}\BibitemShut {NoStop}%
\bibitem [{\citenamefont {Goodenough}\ and\ \citenamefont
  {Manthiram}()}]{Goodenough1991}%
  \BibitemOpen
  \bibfield  {author} {\bibinfo {author} {\bibfnamefont {J.~B.}\ \bibnamefont
  {Goodenough}}\ and\ \bibinfo {author} {\bibfnamefont {A.}~\bibnamefont
  {Manthiram}},\ }\enquote {\bibinfo {title} {Chemistry of high temperature
  superconductors},}\ \BibitemShut {NoStop}%
\bibitem [{\citenamefont {Baughman}\ \emph {et~al.}(1998)\citenamefont
  {Baughman}, \citenamefont {Stafstrom}, \citenamefont {Cui},\ and\
  \citenamefont {Dantas}}]{Baughman1998}%
  \BibitemOpen
  \bibfield  {author} {\bibinfo {author} {\bibfnamefont {R.~H.}\ \bibnamefont
  {Baughman}}, \bibinfo {author} {\bibfnamefont {S.}~\bibnamefont {Stafstrom}},
  \bibinfo {author} {\bibfnamefont {C.}~\bibnamefont {Cui}}, \ and\ \bibinfo
  {author} {\bibfnamefont {S.~O.}\ \bibnamefont {Dantas}},\ }\href@noop {}
  {\bibfield  {journal} {\bibinfo  {journal} {Science}\ }\textbf {\bibinfo
  {volume} {279}},\ \bibinfo {pages} {1522} (\bibinfo {year}
  {1998})}\BibitemShut {NoStop}%
\bibitem [{\citenamefont {Attard}\ \emph {et~al.}(2016)\citenamefont {Attard},
  \citenamefont {Caruana-Gauci}, \citenamefont {Gatt},\ and\ \citenamefont
  {Grima}}]{Attard2016}%
  \BibitemOpen
  \bibfield  {author} {\bibinfo {author} {\bibfnamefont {D.}~\bibnamefont
  {Attard}}, \bibinfo {author} {\bibfnamefont {R.}~\bibnamefont
  {Caruana-Gauci}}, \bibinfo {author} {\bibfnamefont {R.}~\bibnamefont {Gatt}},
  \ and\ \bibinfo {author} {\bibfnamefont {J.~N.}\ \bibnamefont {Grima}},\
  }\href@noop {} {\bibfield  {journal} {\bibinfo  {journal} {Phys. Status
  Solidi B}\ }\textbf {\bibinfo {volume} {253}},\ \bibinfo {pages} {1410}
  (\bibinfo {year} {2016})}\BibitemShut {NoStop}%
\end{thebibliography}

\end{document}